# Multiferroicity and Magnetoelectric Coupling in TbMnO$_3$ Thin Films

*Ni Hu [1,2], Chengliang Lu [3*], Zhengcai Xia [3], Rui Xiong [1], Pengfei Fang [1], Jing Shi [1]\*, and Jun-Ming Liu [4,5]*

[1] Department of Physics, Wuhan University, Wuhan 430072, China

[2] School of Science and Hubei Collaborative Innovation Center for High-Efficiency Utilization of Solar Energy, Hubei University of Technology, Wuhan 430068, China

[3] School of Physics and Wuhan National High Magnetic Field Center, Huazhong University of Science and Technology, Wuhan 430074, China

[4] Laboratory of Solid State Microstructures and Innovation Center of Advanced Microstructures, Nanjing University, Nanjing 210093, China

[5] Institute for Quantum Materials, Hubei Polytechnic University, Huangshi 435000, China

Keywords: Multiferroicity, thin film, magnetoelectric effect, magnetism, dielectric

ABSTRACT: In this work, we report the growth and functional characterizations of multiferroic TbMnO$_3$ thin films grown on Nb-doped SrTiO$_3$ (001) substrates using pulsed laser deposition. By performing detailed magnetic and ferroelectric properties measurements, we demonstrate that the multiferroicity of spin origin known in the bulk crystals can be successfully transferred to TbMnO$_3$ thin films. Meanwhile, anomalous magnetic transition and unusual magnetoelectric coupling related to Tb moments are observed, suggesting a modified magnetic configuration of Tb in the films as compared to the bulk counterpart. In addition, it is found that the magnetoelectric coupling enabled by Tb moments can even be seen far above the Tb spin ordering temperature, which provides a larger temperature range for the magnetoelectric control involving Tb moments.

■ **INTRODUCTION**

Multiferroic manganites $R$MnO$_3$ ($R$ is rare-earth element), hosting intimately coupled ferroelectric (FE) and magnetic orders, are of pressing interest for new multifunctional devices since the discovery of giant magnetoelectric (ME) coupling effect in TbMnO$_3$ (TMO) [1-4]. So far dozens of novel properties have been evidenced in mutlferroic $R$MnO$_3$ thin films and heterostructures, such as emergent ferromagnetism in $R$MnO$_3$ thin films and exchange bias in $R$MnO$_3$ based heterostructures [5-11]. One standard origin of the multiferroicity is the inverse Dzyaloshinskii-Moriya (DM) interaction working between adjacent spins. The as-generated polarization ($P$) can be expressed as $P \sim e_{ij} \times (S_i \times S_j)$, where $e_{ij}$ denotes the unit vector connecting two interacting spins $S_i$ and $S_j$ [12, 13]. Typically in TMO, cooling from high temperature ($T$) paramagnetic phase can drive it into a sinusoidal antiferromagnetic (AFM) state at $T_N \sim 41$ K. Further cooling to $T_C \sim 27$ K, a special spiral-spin-ordering (SSO) arises and an improper spontaneous $P$ emerges along the $c$ axis ($P_c$) *via* the inverse DM mechanism. This induced $P_c$ can be flopped to lie along the $a$ axis ($P_a$) by applying magnetic field in-plane associated with a 90° rotation of the spin spiral plane from the $bc$-plane ($bc$-spiral) to the $ab$-plane ($ab$-spiral) [2, 14].

While TMO presents giant ME coupling related to the magnetic-flop of $P$, the inherent AFM ordering prevents this material to be integrated into memory devices. In fact, this is a common drawback of the whole group of multiferroic materials with spin origin. Recently, a few studies revealed exciting ferromagnetic tendency in TMO thin films through strain and domain engineering [10, 11, 15-17]. In particular, emergent ferromagnetism and anomalous magnetic transition at relatively high temperature (~125 K) were recently observed in TMO thin films grown on (001) SrTiO$_3$ (STO) substrates [16]. These works point to a bright future of improving the ferromagnetism in multiferroic thin films. However, the magnetically-driven FE state as well as its response to magnetic field has not yet been experimentally studied in TMO thin films. More than this, the TMO/STO thin film was supposed to be non-multiferroic, because of the disappearance of $T_C$ in the magnetization data [18, 19]. It seems physically reasonable to anticipate a suppression of the SSO by strain in thin films, since this special magnetic order just exists in a pretty narrow region in the phase diagram of $R$MnO$_3$ [20]. Nevertheless, detailed characterizations are still desired to give any conclusive answer to the multiferroicity and ME coupling in TMO thin films. Moreover, although the multiferroicity and ME control in bulk crystals have been intensively investigated in the last decade, the related research in multiferroic thin films is still rare, and the insights would be significant in designing advanced multifunctional devices.

Recently, by performing ultra fast optical spectroscopy on TMO/STO (001) thin films, Qi *et al.* revealed an anomalous decay-time-constant at ~30 K which is close to the $T_C$ ~27 K in bulk TMO, indicating the existence of coupled FE and magnetic orders in the films [21]. Moreover, a recent work observed the ferroelectric state in TMO thin films deposited on the top of YAlO$_3$ (100) substrates by utilizing optical second harmonic generation. In particular, this work further found that the noncollinear magnetic state could exist in thin films with thickness down to 6 nm, implying that the SSO in TMO is kind of robust against (modest) strain and spatial confinement [22]. These findings indeed motivate us to explore the magnetic multiferroicity and ME coupling in TMO thin films.

In the present work, we address this issue by performing systematic magnetic and ferroelectric measurements in TMO/STO (001) thin films. Our results demonstrate that the multiferroic properties seen in the TMO bulk crystals can be successfully preserved in thin films. Besides, the strain still performs certain modification on the ME coupling in the films, which can be associated with the Tb moments.

■ **EXPERIMENTAL DETAILS**

Epitaxial TMO thin films with thickness $t$=270 nm were grown on (001) Nb-doped SrTiO$_3$ (NSTO) single crystal substrates using pulsed laser deposition (PLD) method. The deposition was carried out at 800 °C and 0.2 mbar oxygen partial pressure. A KrF excimer laser ($\lambda$=248 nm) with an energy density of ~1 J/cm$^2$ was used for the ablation. A stoichiometric TMO ceramic target was used for the deposition. After deposition, the thin films were cooled down at a rate of 5 °C/min under a 400 mbar oxygen ambient to achieve a full oxidization. High resolution x-ray diffraction (HRXRD) was utilized to check the pure phase of the thin films. Scanning Electron Microscopy (SEM) was used to determine the thickness of the films. Magnetization (*M*) as a function of *T* and magnetic field (*H*) was measured using a Magnetic Properties Measurement System (MPMS, Quantum Design). The *M(T)* curves were measured for both field cooling (FC) and zero field cooling (ZFC) sequences, and the cooling and measuring field was set at *H*=0.1 T. All the *M(H)* curves were measured after ZFC. In order to detect the ferroelectric polarization, a pyroelectric current method was utilized. Detailed measurements of dielectric permeability of the *c* axis $\varepsilon_c$ as a function of *T* and *H* were performed using a LCR meter which connects to a Physical Properties Measurement System (PPMS, Quantum Design).

■ **RESULTS AND DISCUSSION**

Figure 1 presents the HRXRD pattern of one typical TMO/NSTO (001) thin film, which confirms the pure phase and *c*-axis orientation growth of the thin film. The out-of-plane lattice parameter of the film can be derived to be *c*=7.416 Å which is slightly larger than the bulk value (7.403 Å), evidencing a modest in-plane compressive strain. This has previously been demonstrated to arise from strain relaxation by development of an in-plane twin-like crystalline structure [18, 19]. Since the present TbMnO$_3$ thin films only have weak strain, we expect that the magnetically induced ferroelectricity known in the strain-free TbMnO$_3$ can be preserved in the present thin films. As revealed by earlier structural investigations, the lattice parameters of TMO/STO (001) thin films are almost independent on the film thickness when $t >20$ nm [18]. Moreover, multiferroicity known in the bulk TMO is still missed in the corresponding thin films even when the film thickness goes to ~200 nm [10, 16]. Therefore, to enable the physical property measurements such as magnetization, the film thickness was set at a relatively large value of $t=270$ nm in the present work, which can be determined by cross-section SEM characterization shown in the inset of Fig. 1.

As shown in Fig. 2(a), the *T*-dependence of *M* measured along the *c* axis and in the *ab* plane show smooth variation down to $T \sim T_R \sim 9$ K, at which peak anomalies emerge due to the magnetic ordering of Tb moments. This is different from the observation in TbMnO$_3$ bulk crystals, in which clear AFM transition of Mn is seen in *M(T)* curves [2]. The large paramagnetic contribution of Tb moments could be one of the possible origins responsible for such discrepancy, which will be discussed in the following. Here we note that the shape of *M(T)* curves show clear difference, indicating anisotropic magnetism in the thin films. This is further confirmed by measurements of *M(H)*, as shown in Fig. 2(b)-(d). As $H>2$ T and below $T_R$, the *M(H)* curves of both directions show saturation, and the ratio of saturated *M* of the two directions is $M_{ab}/M_c \sim 3$, suggesting the hard *c* axis. The unsaturated *M(H)* curves as $T>T_R$ should be ascribed to the paramagnetic contribution from Tb.

Here, it is worth mentioning that a quick enhancement of *M* is seen at $H\sim1.5$ T in the *M(H)* curve measured at $T=2$ K with *H//ab* plane, shown in Fig. 2(b). In fact, such magnetic anomaly (indicated by olive arrows) can be observed for both *H//ab* plane and *H//c* axis in the present thin films, as shown in Fig. 2(c) and (d). However, in bulk TMO, it has been identified that such phenomenon can only be triggered by the application of magnetic field in the *ab*-plane, arising from the *H*-induced spin-reorientation of Tb [14]. This discrepancy suggests a modified magnetic sublattice of Tb in the films. In addition, the emergence of such magnetic transition along the *c* axis suggests an enhanced out-of-plane component of the Tb moments. This is consistent with the reduced magnetic anisotropy ($M_{ab}/M_c \sim 3$) in the thin films as compared to values in the bulk counterpart ($M_a/M_c \sim 6$, $M_b/M_c \sim 4$) [14].

In order to detect possible magnetically induced multiferroicity in the thin films, we performed detailed dielectric permeability $\varepsilon_c$ measurements with *H* applied along the *c* axis and in the *ab* plane. Fig. 3 (a) and (b) presents measured $\varepsilon_c$ as a function of *T* under various magnetic fields. Upon cooling at *H*=0 T, a sharp cusp emerges at $T=T_C \sim 26.5$K. Correspondingly, spontaneous electric polarization arises at the same point, shown in Fig. 3(d). These demonstrate the occurrence of ferroelectricity in the present thin films, arising from the development of Mn-SSO via the inverse DM interaction similar to the case in bulk TMO [2]. The measured electric polarization is switchable and intrinsic, as evidenced by performing measurements with different poling-field and warming rate, shown in Fig. 3(c) and (d). According to the phase diagram of bulk TMO [14, 23], the $T_C$ of the magnetically induced FE state presents anisotropic response to *H*, which can be understood as following. As *H*//*c* axis in TMO, the magnetic field acts as a role of destructing the Mn-SSO, leading to a shift of $T_C$ towards low-*T*. However, the application of *H* in the *ab*-plane would rather flop the Mn-SSO from the *bc*-plane to the *ab*-plane than suppress the spin spiral phase, resulting in quite stable $T_C$ against *H*. Indeed, such anisotropic response of $T_C$ to *H* is well reproduced in the present thin films, as shown in Fig. 3(c), confirming the similar multiferroic physics for both the TMO films and its bulk counterpart.

While the above data demonstrate that the multiferroicity known in the bulk TMO has been successfully transferred into the present thin films, still some different points can be identified. For instance, magnetic anomaly in $M(H)$ is observed for both $H//c$ axis and $H//ab$ plane in the thin films, but it only happens for $H//ab$ plane in the bulk [14]. This may promise an unusual ME coupling, arising from the strong Tb-Mn spin interaction in TMO. Fig. 4(a) displays $H$ dependence of $\varepsilon_c$ measured at 5 K, in which two sharp peaks arise at $H_{C1} \sim 1.5$ T and $H_{C2} \sim 5.1$ T, respectively. Correspondingly, the measured dielectric loss tan$\delta$ as a function of $H$ also shows two clear peak anomalies at the two fields, as shown in Fig. 4(b). For the first peak anomaly, the critical field $H_{C1} \sim 1.5$ T perfectly matches the magnetic transition in $M(H)$ (Fig. 2(d)), verifying the important role of Tb moments in mediating ME coupling in TMO. Such a direct correspondence between magnetic transition and dielectric anomaly can be unconventionally observed for $H//c$ axis up to $T_R$, as shown in Fig. 4(c). Moreover, although $H_{C1}$ of $H//ab$ plane, indicated by olive bars in Fig. 4(d), gradually shifts to low-$H$ region upon increasing $T$, it still can be observed even up to 20 K which is far higher than $T_R \sim 9$ K. These phenomena related to Tb moments are quite different from those seen in the bulk, pointing to a modified role of Tb moments in tuning the ME effect in TMO thin films.

With regard to the peak anomaly appearing at $H_{C2} \sim 5.1$ T, it is quite close to the critical field of the $H$-driven $P$-flop transition ($H_{flop} \sim 4.6$ T) in TMO bulk crystals [14]. Moreover, it is note that $H_{C2}$ obtained at various temperatures in the films are all close to the values of $H_{flop}$ obtained in the bulk crystals, and both $H_{C2}$ and $H_{flop}$ evolves with $T$ coherently, shown in Fig. 5. Therefore, the peak anomaly at $H_{C2} \sim 5.1$ T seen in the present films is probably also due to a $H$-induced $P$-flop transition, similar to the case in the bulk counterpart. Actually, the $H$-driven $P$-flop is a general property of multiferroic $R$MnO$_3$ with SSO, and such a transition is usually accompanied by evident hysteresis when sweeping $H$ up and down [14, 24]. In the present TMO thin films, striking hysteresis around $H_{C2}$ can be seen in both $\varepsilon_c$-$H$ and $\tan\delta$-$H$ curves (Fig. 4(a) and (b)), agreeing with a possible $P$-flop transition induced by the application of magnetic field. Nevertheless, direct evidence that in-plane electric polarization occurs above $H_{C2}$ is desired to demonstrate a $H$-induced $P$-flop transition, while this is unfortunately not feasible for the present experimental geometry.

In the phase diagram summarized in Fig. 5, it is seen that $H_{C1}$ is intriguingly observed even above $T_R$. Simultaneously, in Fig. 2, magnetic transitions in $M(H)$ are unconventionally observed for both $H//c$ and $H//ab$ plane. These phenomena enabled by Tb moments are quite different from the bulk, suggesting a modified magnetic Tb sublattice in the present thin films. However, the FE characterizations shown above clearly demonstrate that the multiferroicity known in the bulk has been preserved in the present TMO films, evidencing "copied" Mn-SSO in the films. Therefore, it would be natural to assign the disappearance of the AFM transition in the present $M(T)$ curves to the fluctuation of Tb. In fact, it is noted that the fluctuation of the large $4f$ moment of $R$ contributes to magnetization significantly above $T_R$, which could blur the AFM transition of Mn in multiferroics $R$MnO$_3$. For instance in orthorhombic YMnO$_3$ and HoMnO$_3$, the two materials are quite similar in terms of crystalline and magnetic structures except that $Y^{3+}$ is nonmagnetic but Ho$^{3+}$ has large magnetic moment. However, clear AFM transition is seen in the $M(T)$ for YMnO$_3$ single crystals [25], thin films [26, 27], and polycrystals [28], but no AFM transition can be deduced from the $M(T)$ for HoMnO$_3$ [29]. In orthorhombic YMnO$_3$ and its derivates with nonmagnetic $R$ [30-32], the AFM transition was commonly observed in $M(T)$, while this is challenge for multiferroic $R$MnO$_3$ with magnetic $R$ such as DyMnO$_3$ [33] and HoMnO$_3$ [29]. Nevertheless, more microscopic information are certainly desired to clarify this issue.

As revealed by a few recent theoretical and experimental studies, the spin interaction between $R$ and Mn plays a significant role in tuning the multiferroicity in $R$MnO$_3$, such as the huge reduction of $P$-flop critical filed after involving the Tb moments [23, 29, 34-38]. In particular, it was revealed that such $R$-Mn spin interaction could remarkably enhance the magnetically driven FE polarization *via* a so-called exchange striction effect, and thus the related ME coupling could be magnified significantly, typically in DyMnO$_3$ [35, 37]. However, the large ME control related to $R$-Mn spin interaction is usually only accessible below $T_R$ in these materials. In the present TMO thin films with modest strain, the ME coupling arising from Tb-Mn interaction can be extended even up to $T_C > T_R$. This promises a larger temperature range for manipulating the ME effect enabled by the Tb-Mn spin interaction, and would have important implication of exploring unusual ME control facilitated by $R$-Mn coupling in other multiferroic $R$MnO$_3$ thin films with strain engineering.

■ **CONCLUSION**

In summary, multiferroic TbMnO$_3$ thin films with pure *c*-orientation have been successfully synthesized on the top of Nb-doped SrTiO$_3$ (001) substrates using pulsed laser deposition, and the physical properties, including magnetism, ferroelectricity, and magnetoelectric coupling have been investigated in detail. Our results revealed that the TbMnO$_3$ films possess magnetically driven multiferroicity similar to the case in the bulk counterpart. However, different magnetic properties and magnetoelectric coupling associated with Tb moments were observed, indicating modified magnetic sublattice of Tb in the films. This is further supported by the observation of unusual magnetoelectric coupling enabled by Tb moments in the *ab*-plane above $T_R$.

■ **AUTHOR INFORMATION**

**Corresponding Author**

*Authors to whom correspondence should be addressed, email: cllu@mail.hust.edu.cn;

jshi@whu.edu.cn

**Author Contributions**

N.H. and C.L.L. designed and performed the experiments. Z.C.X. did the magnetic measurements. N.H., C.L.L., J.S., and J.M.L. wrote the manuscript. Z.C.X., R.X., and P.F.F. contributed to the detailed discussions and revisions.

**Notes**

The authors declare no competing financial interest.

■ ACKNOWLEDGMENT

This work was supported by the National Nature Science Foundation of China (Grant Nos. 11374112, 11374147, and 11304091), the National 973 Project of China (Grant No. 2011CB922101). C.L.L. acknowledges the funding from the Alexander von Humboldt Foundation.

■ FIGURES

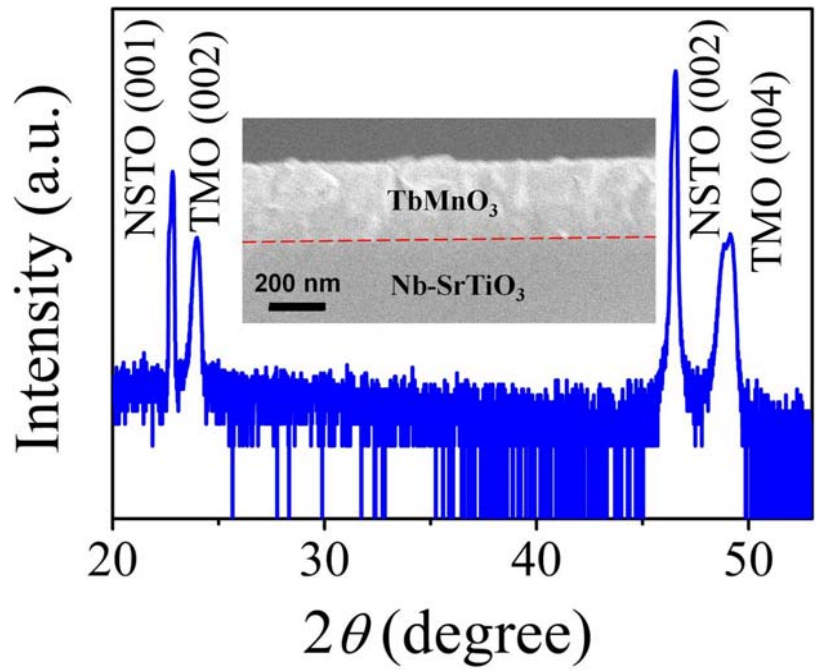

**Figure 1.** X-ray diffraction $\theta$-$2\theta$ scan of a TbMnO$_3$/Nb-SrTiO$_3$ (001) thin film. The inset shows a cross-section scanning electron microscopy of the film, from which the film thickness is derived to be ~270 nm.

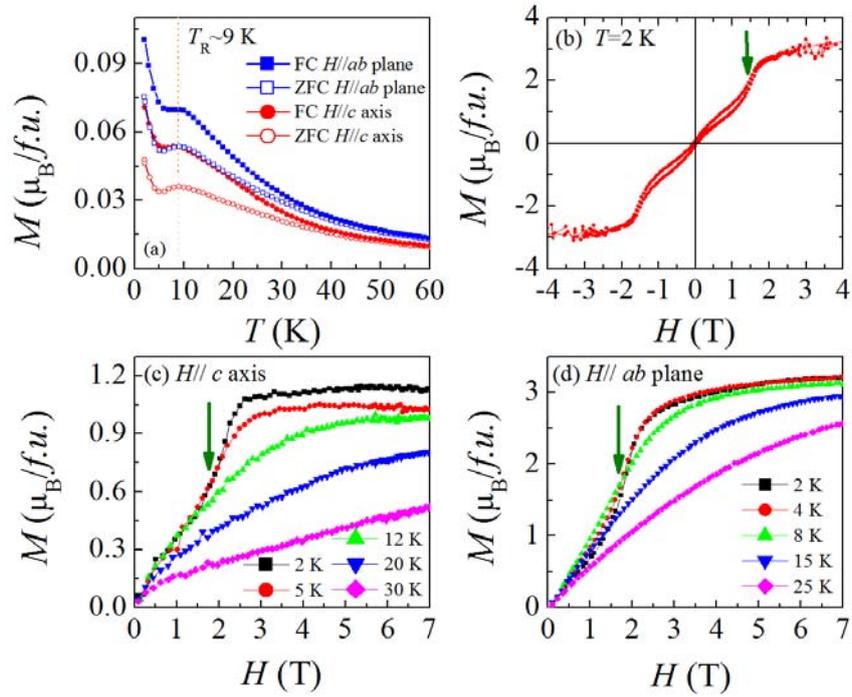

**Figure 2.** (a) Magnetization as a function of temperature with magnetic field applied along the *c* axis and in the *ab*-plane. (b) Magnetic hysteresis loop with *H//ab* plane measured at *T*=2 K. Magnetic field dependence of magnetization measured at various temperatures with (c) *H//c* axis, and (d) *H//ab* plane.

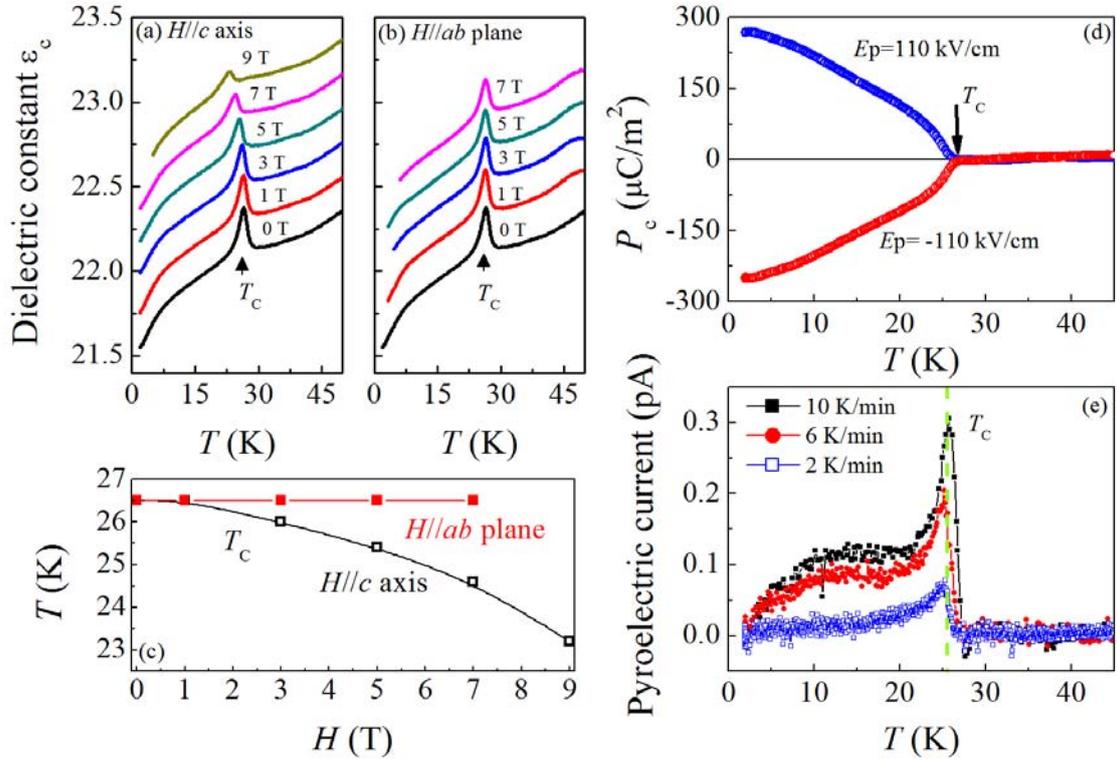

**Figure 3.** Dielectric constant $\varepsilon_c$ as a function of temperature measured under various magnetic fields applied along (a) the *c* axis, (b) and in the *ab*-plane. The phase transitions are indicated by arrows. (c) Evaluated $T_C$ as a function of magnetic field of *H//c* axis and *H//ab*-plane. (d) Measured spontaneous polarization as a function of temperature after poling in different electric fields $E_p$, in which switchable polarization can be observed. (e) Collected pyroelectric current as a function of temperature with different warming rates.

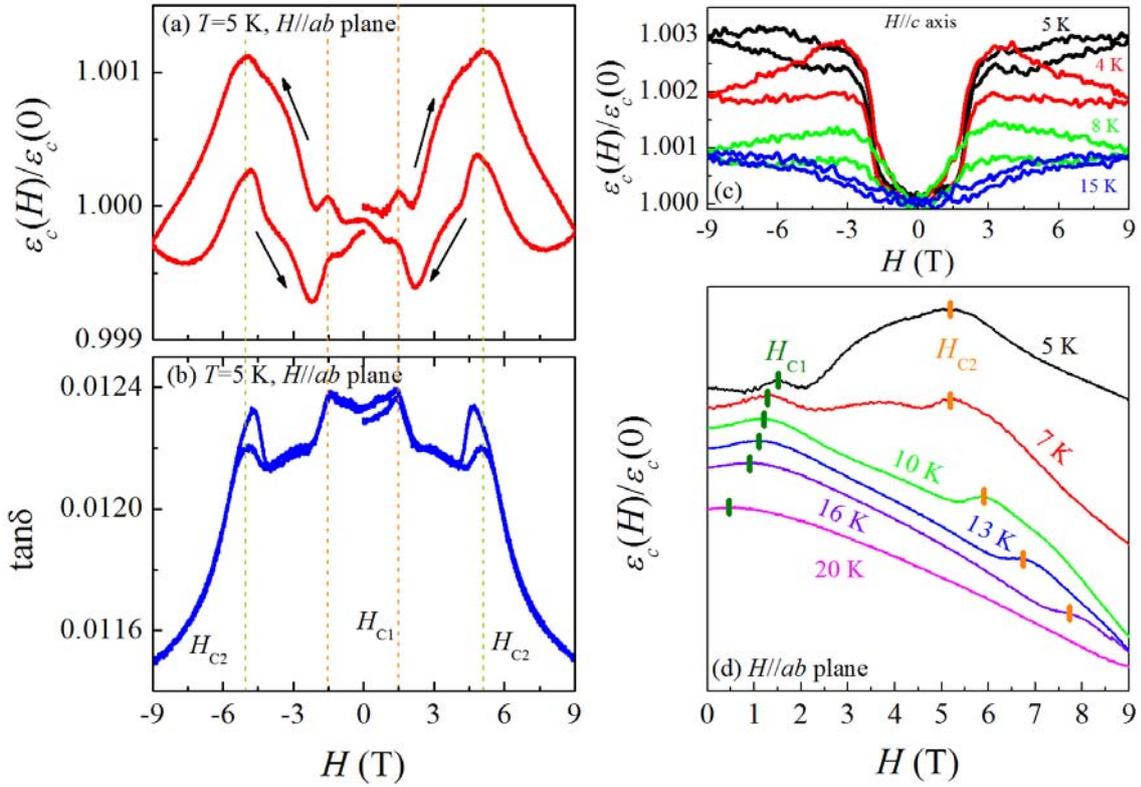

**Figure 4.** (a) Normalized dielectric constant $\varepsilon_c(H)/\varepsilon_c(0)$ as a function of magnetic field applied in the *ab*-plane at 5 K. (b) The corresponding dielectric loss tanδ vs *H*. (c) *H* dependence of normalized $\varepsilon_c(H)/\varepsilon_c(0)$ measured at various temperatures with *H*//*c* axis. (d) *H* dependence of normalized $\varepsilon_c(H)/\varepsilon_c(0)$ measured at various temperatures with *H*//*ab*-plane.

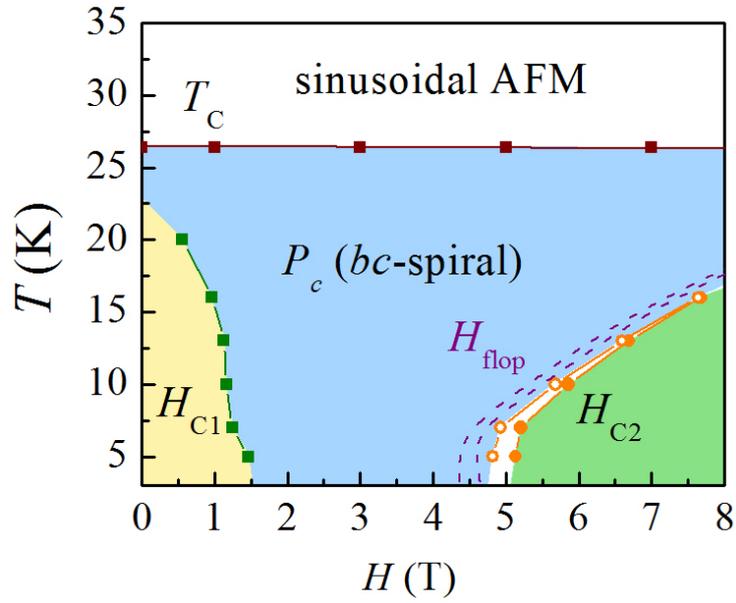

**Figure 5.** Multiferroic phase diagram of the TbMnO$_3$ thin films with $H//ab$-plane. For a comparison, the critical fields of $P$-flop transition ($H_{flop}$) in TbMnO$_3$ bulk crystals are also plotted (dashed purple curves), which are quite close to $H_{C2}$.

**Table of Contents Graphic**

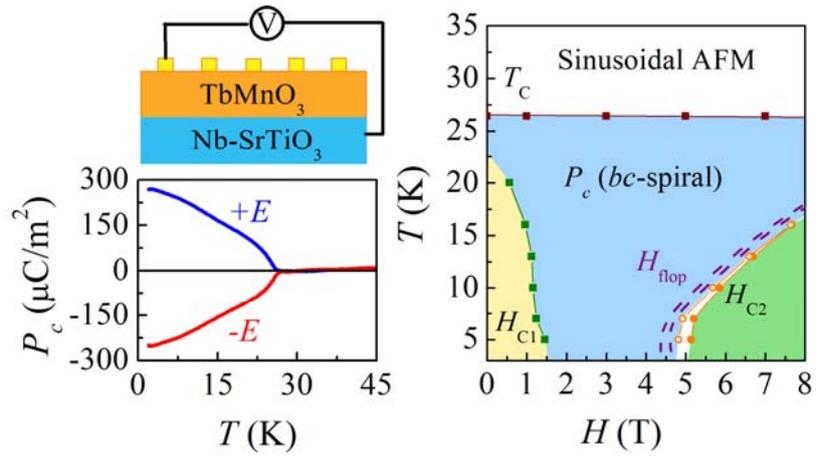